# Revealing the doping density in perovskite solar cells and its impact on device performance

Francisco Peña-Camargo[1], Jarla Thiesbrummel[1,2], Hannes Hempel[3], Artem Musiienko[4], Vincent M. Le Corre[1,5], Jonas Diekmann[1], Jonathan Warby[1], Thomas Unold[3], Felix Lang[1], Dieter Neher[1], Martin Stolterfoht[1,*]

1. Physik weicher Materie, Institut für Physik und Astronomie, Universität Potsdam, Karl-Liebknecht-Str. 24–25, 14776 Potsdam, Germany
2. Clarendon Laboratory, University of Oxford, Parks Road, Oxford OX1 3PU, United Kingdom
3. Department of Structure and Dynamics of Energy Materials, Helmholtz-Zentrum Berlin für Materialien und Energie, D-14109, Berlin, Germany
4. Institut für Silizium-Photovoltaik, Helmholtz-Zentrum Berlin für Materialien und Energie, Kekuléstrasse 5, 12489, Berlin, Germany
5. Institute of Materials for Electronics and Energy Technology, Friedrich-Alexander-Universität Erlangen-Nürnberg, 91058, Erlangen, Germany

*email: stolterf@uni-potsdam.de

**Abstract**

Traditional inorganic semiconductors can be electronically doped with high precision. Conversely, there is still conjecture regarding the assessment of the electronic doping density in metal-halide perovskites, not to mention of a control thereof. This paper presents a multifaceted approach to determine the electronic doping density for a range of different lead-halide perovskite systems. Optical and electrical characterisation techniques comprising intensity-dependent and transient photoluminescence, AC Hall effect, transfer-length-methods, and charge extraction measurements were instrumental in quantifying an upper limit for the doping density. The obtained values are subsequently compared to the charge on the electrodes per unit volume at short-circuit conditions, which amounts to roughly $10^{16}$ cm$^{-3}$. This figure equals the product of the capacitance $C$ and the built-in potential $V_{\text{bi}}$ and represents the critical limit below which doping-induced charges do not influence the device performance. The experimental results demonstrate consistently that the doping density is below this critical threshold (< $10^{12}$ cm$^{-3}$ which means < $CV_{\text{bi}}$ per unit volume) for all common lead-based metal-halide perovskites. Nevertheless, although the density of doping-induced charges is too low to redistribute the built-in voltage in the perovskite active layer, mobile ions are present in sufficient quantities to create space-charge-regions in the active layer, reminiscent of doped *pn*-junctions. These results are well supported by drift-diffusion simulations which confirm that the device performance is not affected by such low doping densities.

**Introduction**

Traditionally, many inorganic semiconductor devices such as transistors, photodetectors, light-emitting diodes, and solar cells, among others, are based on electronic junctions created by electrical doping. Doping of semiconductors is generally achieved by the introduction of atoms with higher (lower) valency than the substituted atoms, which creates free electrons (holes) in the conduction (valence) band, thereby increasing (lowering) the Fermi level of the semiconductor. The dark concentration of carriers determines the position of the Fermi level. For substitutional doping in inorganic semiconductors, dopants are, for example, introduced by vapour-phase epitaxy. This allows to tune their concentration with high precision, reaching concentrations up to $10^{20}$ cm$^{-3}$ for GaAs (ref.[1]), Ge, crystalline Si (ref.[1]) and amorphous Si (ref.[3]). Substitutional doping introduces ionised shallow defects[2], which can be *intrinsic* (vacancies, interstitials, or substitutions) or *extrinsic* (impurities). In solution-processed semiconductors such as organic semiconductors, on the other hand, dopants can be introduced by adding molecules with very-low-lying lowest unoccupied molecular orbital (LUMO) energy levels for *p*-doping or, conversely, very-high-lying highest occupied molecular orbital (HOMO) energy levels for *n*-doping.[4–8]

In contrast to crystalline inorganic semiconductors, doping of the absorber layer is a more controversial topic in *perovskites*. For example, the community refers to perovskite solar cells as having a *pin* or *nip* architecture, which underlines that the active layer is expected to be intrinsic, i.e. the concentration of the free charges present in the semiconductor at thermal equilibrium is so low that the Fermi level remains in the middle of the bandgap.[9] On the other hand, several publications suggested that perovskites are strongly doped, with reported densities as high as $10^{20}$ cm$^{-3}$ (ref.[10–14]). Moreover, it has even been suggested that the working mechanism of perovskite solar cells is based on the formation of *pn*-junctions in the absorber layer, reminiscent of Si solar cells.[15] The controversy with regards to whether lead-halide perovskites behave as intrinsic semiconductors or not, is a result of various aspects:

a) *Doping* is often referred to as a mere addition of certain additives to the precursor solutions. Although additives may dope a material in a chemical sense, this does not mean the material is also electronically doped. The use of the word '*doping*' without clarifying its type, contributes to confusion. From here on, we will refer to *doping density* as the concentration of free majority charge carriers in the neat bulk material in the dark at room temperature.

b) Electronic doping can be achieved by the introduction of intrinsic defects in the APbX$_3$ lattice[9]. The presence of both donor and acceptor shallow defects was previously demonstrated in MAPbI$_3$ single crystals.[16] Density Functional Theory (DFT) calculations show the possibility of tailoring doping densities as a consequence of defect formation.[17] Changing the ratio PbX$_2$/AX in the fabrication procedure can induce intrinsic defects, and some studies reported doping densities in the order of $10^{18}$ cm$^{-3}$ achieved in this way.[14] However, further experimental evidences point to a general limitation for carrier densities of MAPbI$_3$ at <$10^{14}$ cm$^{-3}$ from both stoichiometric and non-stoichiometric conditions.[15,18,19]

c) Photoemission spectroscopy measurements often suggest an apparent *p*- or *n*-type perovskite layer. However, charge transfer and charge redistribution give rise to an energetic alignment of the Fermi level of the perovskite and the substrate, causing a shift of the energy levels and associated band bending.[20] Consequently, depending on the substrate, different conclusions might be obtained with respect to the doping of the perovskite layer, as was recently shown by Shin, D. et al.[21] Furthermore, although several groups have observed an *n*-type surface, this does not imply that the bulk is doped, as ultraviolet photoemission spectroscopy measures only the energetics at the surface.[22–24] To quantify the band bending within the perovskite absorber layer or the bottom interface, surface photovoltage (SPV) measurements can be performed using white background light. Upon saturation of the valence band onset with light intensity, the bands are flattened, thereby revealing the energy levels of the bulk. Indeed, for a triple cation perovskite $Cs_{0.05}(FA_{0.83}MA_{0.17})_{0.95}Pb(I_{0.83}Br_{0.17})_3$ on a PTAA/PFN-Br substrate, these measurements have demonstrated that the Fermi level in the perovskite bulk is close to midgap[22–24] (see **Supporting Information Figure S1** for further details). However, in several publications on photoemission spectroscopy on perovskites, where the perovskites was found to be highly doped (e.g. ref.[25]), the numerical value of the doping density was not specified. This information is necessary from the device perspective as there is a critical threshold above which the doping density starts to impact the device performance.

d) High concentrations of mobile ions between $10^{15}$ and $10^{19}$ cm$^{-3}$ have been reported for lead halide perovskites[26–30] and ion-induced field-redistribution could be misinterpreted as redistribution of doping-induced electrons or holes.

e) Mott-Schottky (MS) plots are widely used to evaluate the doping density. However, the interpretation of the data is not straightforward. In particular, the MS analysis has been developed for doped semiconductors in the bias region where the depletion width *w* is smaller than the device thickness *d* (refs [31–33]). For perovskite systems with an active layer thickness of 500 nm and a built-in potential $V_{bi}$ of 1 V (ref.[34]), this implies that at least $10^{16}$ cm$^{-3}$ free charges must be present in the device to evaluate the doping density using the MS analysis. Another equivalent definition at which the doping density matters can be introduced by considering the charge on the electrodes, which will be discussed in this manuscript.

Generally, the doping density can be assessed directly or with techniques that require complementary characterisation, as recently summarised by Euvard, et al.[9] Techniques like four-point-probe,[35] transfer-length method,[36] and Van der Pauw,[37] are used to evaluate the doping density via the electrical conductivity, assuming a constant mobility.[38] The assessment of the doping density is also possible using Hall effect measurements and Mott-Schottky analysis. However, as mentioned previously, the interpretation of the latter is not straightforward, especially in low-doped and very thin films, where the doping of the active layer cannot be disentangled from injected charges from the contacts and recombination.[31,39] For semiconductors with low mobility (≤ 1 cm$^2$ V$^{-1}$ s$^{-1}$) and high resistivity (≥ $10^5$ Ω cm), Hall effect has to be performed in an alternative way to the static configuration. Lock-in detection and AC magnetic fields are used to overcome this issue.[16,18,40] In addition, doping can be also investigated indirectly, for example via intensity dependent photoluminescence quantum efficiency[41] (PLQY) or via pump-probe spectroscopy by means of i.e. transient photoluminescence (TRPL)[42] and time-resolved microwave conductivity.[43] Charge extraction techniques such as Bias-Assisted Charge Extraction (BACE)[44], space-charged-limited currents[45], Charge Extraction under Linearly Increasing Voltage (CELIV)[46], are also known to be suitable for the assessment of the doping density of organic,[32] inorganic[46] and perovskite solar cells.[47] We also note that some methods, such as PLQY, charge extraction, and Mott-Schottky, are more limited than Hall effect measurements in terms of sensitivity, i.e. it may only possible to obtain upper limits of the doping density if the doping concentrations are very low.

In this study, we investigate and quantify the electronic doping density for a range of six different sorts of lead-halide perovskites by several highly sensitive and reliable methods. To obtain the free carrier density in the dark, Hall effect measurements were performed in a four-contact van der Pauw square arrangement, with an alternating current (AC) magnetic field (50 mHz) and a lock-in amplification to enhance the Hall signal. The determined free carrier densities are of the order of $5\times10^{11}$ - $3\times10^{13}$ cm$^{-3}$. Furthermore, we used transistor-like structures to evaluate the conductivity of thin films through the transfer length method in the dark. In the corresponding neat films, we measured the intensity-dependent PLQY to compare the trend of the experimental data with a recombination model. This approach yields an upper limit for the doping-induced charge density of the order of $10^{12}$ cm$^{-3}$. These results are further confirmed by TRPL measurements, which also demonstrate a density below $10^{12}$ cm$^{-3}$ for Pb-based perovskites. We then performed CELIV measurements in the dark to investigate the doping in complete devices. In addition, intensity-dependent photo-CELIV was measured to demonstrate the sensitivity of the CELIV method and to support the conclusions drawn from these measurements. Finally, we corroborated these results with drift-diffusion simulations to demonstrate that the device performance is not affected by the measured densities of doping-induced charge.

**Results**

To generalise our findings, our investigated set of perovskites is comprised of the archetypal methylammonium lead iodide MAPbI$_3$, a double cation/mixed halide Cs$_{0.15}$FA$_{0.85}$Pb(I$_{0.75}$Br$_{0.25}$)$_3$, and four different triple cation Cs$_{0.05}$(FA$_x$MA$_y$)$_{0.95}$Pb(I$_x$Br$_y$)$_3$ where the percentages of formamidinium lead iodide FAPbI$_3$ ($x$) and methylammonium lead bromide MAPbBr$_3$($y$) were varied. For all these compositions, thin neat films, transistor-like structures, and full devices were fabricated. Besides, neat perovskite samples, in the van der Pauw configuration with gold contacts, were prepared for Hall Effect measurements. The structure of the full devices follows a *pin*-architecture comprising ITO (150 nm)/PTAA:PFN-Br (8 nm)/perovskite (400 nm)/C$_{60}$ (30 nm)/BCP (8 nm)/Cu (100 nm)], where ITO is indium-doped tin oxide, PTAA is poly[bis(4-phenyl)(2,4,6-trimethylphenyl)amine], PFN-Br is poly[(9,9-bis(30-((N,N-dimethyl)-N-ethylammonium)-propyl)-2,7-fluorene)-alt-2,7-(9,9-dioctylfluorene)] dibromide, and BCP is bathocuproine (further details are provided in the **Supporting Information Methods**). Their corresponding *JV* curves are shown in **Supporting Information Figure S2**. A summary of the photovoltaic parameters is also shown in **Supporting Information Figure S3**. Before introducing the experimental methods to quantify the doping density, we discuss at which densities electronic doping becomes relevant for solar cell performance.

When does the doping density matter?

Kirchartz and Cahen have recently discussed that for a *pn*-junction to be formed, the doping density should be above approximately $10^{16}$ cm$^{-3}$ (ref.[48]). In addition to this explanation, a simple threshold at which the doping density becomes relevant for the device performance can be obtained as follows.

Let $C = \epsilon_0\epsilon_r S/d$ be the geometrical capacitance of a solar cell, where $S$ is its area, $\epsilon_0$, the permittivity of free space and $\epsilon_r$, the relative dielectric constant of the absorber layer. When such a cell is short-circuited, a charge equivalent to $CV_\text{bi}$ will be accumulated on its electrodes. This quantity, divided by the cell volume, with an active layer thickness of 500 nm, is of the order of $10^{16}$ cm$^{-3}$ for perovskite systems.[48,49] As long as their density is significantly below the $CV_\text{bi}$ charge per unit volume, the impact of doping-induced equilibrium charges on the internal built-in field will be negligible. Consequently, the device operation would not be affected. The results of drift-diffusion simulations, displayed in **Figure 1**, are consistent with these basic considerations. These simulations, which have been tailored to describe efficient triple cation perovskite solar cells, as discuss in our previous work,[34] demonstrate that the doping-induced charges will only redistribute the internal field at densities higher than $10^{16}$ cm$^{-3}$. Moreover, this rule of thumb holds regardless of the type of doping, whether it is *n* or *p*. The corresponding *JV*s for a range of donor ($n_0$) doping densities are shown in **Figure 1a,** whereas those for a range of acceptor densities ($p_0$) are shown in **Figure 1b**. In both the cases, the device performance is impacted when $n_0, p_0 \geq 10^{16}$ cm$^{-3}$. To show that increasing the doping density beyond $CV_\text{bi}$ yields a redistribution of the internal field, **Figure 1c** represents the band diagram at a device bias of 0 V for some acceptor doping densities, simulated with the same parameters used in **Figure 1b**. As a conclusive criterion, the $CV_\text{bi}$ charge per unit volume marks the critical density at which doping matters – a concept that is generally valid for solar cells. Notwithstanding these points, it is an interesting question, whether the presence of mobile ions in perovskites, with densities reported in literature ranging from $10^{15}$ cm$^{-3}$ to $10^{20}$ cm$^{-3}$ (references[27–30]), can influence this criterion. To check this statement, we performed additional simulations, with a mobile ion density of $10^{17}$ cm$^{-3}$ (either mobile cations or anions, while immobilising the counterpart) and variable doping densities (**Supporting Information Figure S6**). We found that the device performance still only deviates from the intrinsic case at a doping density of $10^{16}$ cm$^{-3}$, i.e. the $CV_\text{bi}$ criterion holds even in the presence of mobile ions. We note that we will discuss the importance of ionic charges with respect to doping induced charges further below.

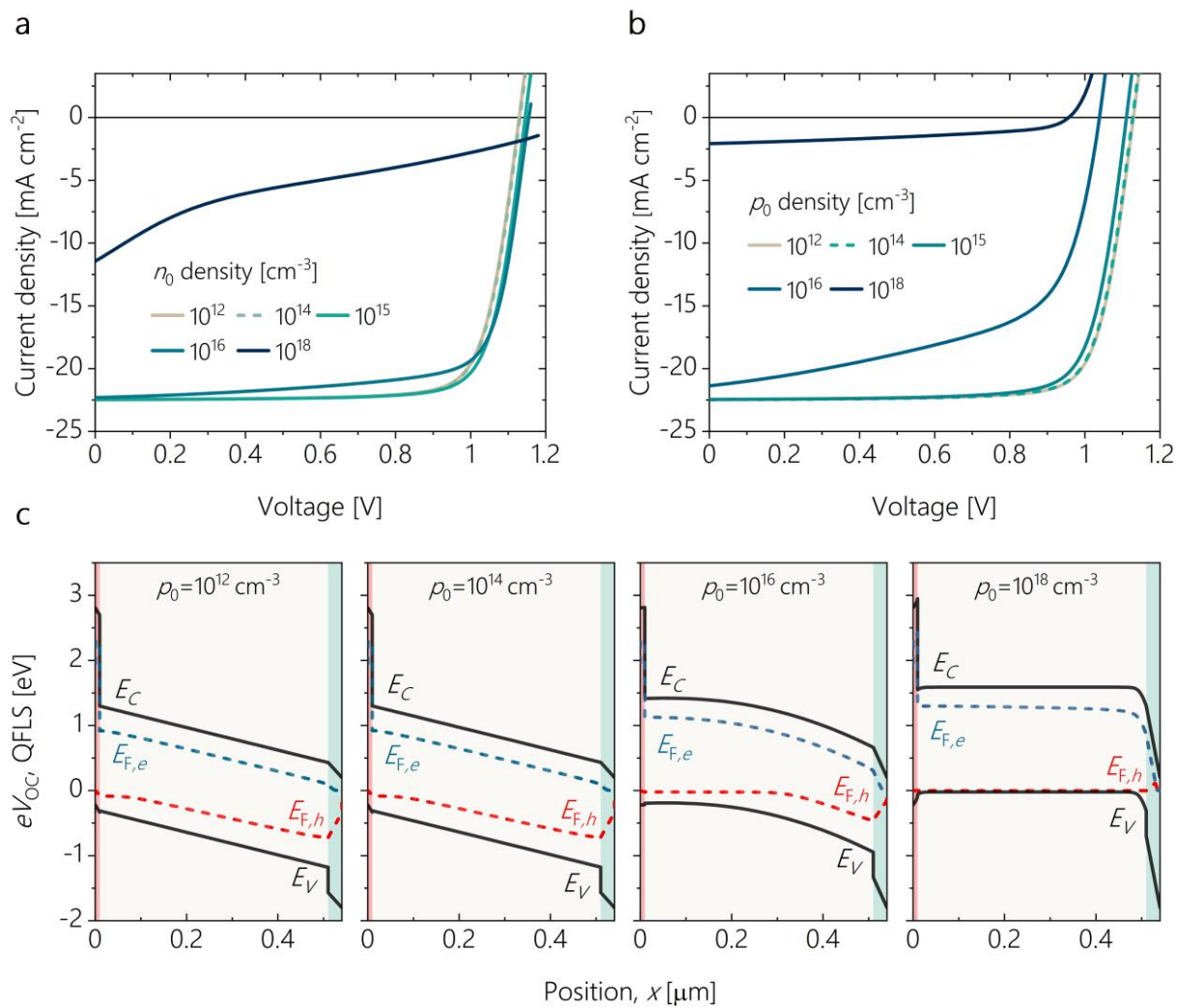

**Figure 1.** Current density versus voltage curves simulated using SCAPS[50,51] without mobile ions as a function of the doping by **a**, donors ($n_0$) and **b**, acceptors ($p_0$). Below the $CV_{bi}$ charge per unit volume (approx. $10^{16}$ cm$^{-3}$), the doping density has no influence on the device performance. **c**, Corresponding band diagrams at zero voltage for various acceptor doping densities without mobile ions. Simulations with mobile ions can be found in the **Supporting Information Figure S6**.

Hall Effect, transfer length method, and surface conductivity

Hall effect measurements enable us to determine the carrier concentration by means of probing the interaction of the charge carriers with an applied magnetic field. Samples were prepared in the traditional van der Pauw layout[37] with four gold contacts placed at the corners of a square-shaped perovskite film.

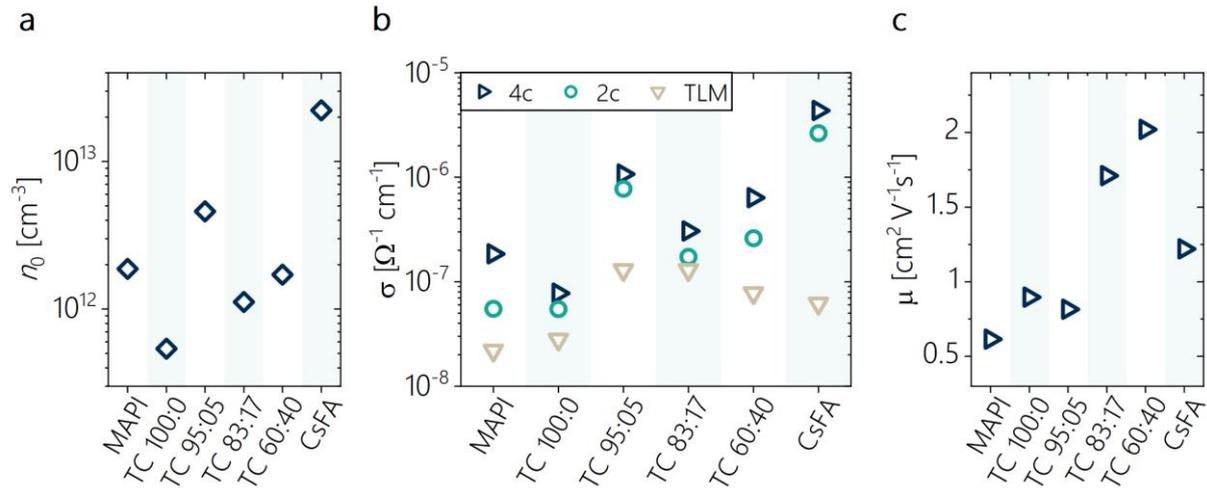

**Figure 2**. **a**, Free charge carrier density in the dark $n_0$ determined by AC Hall effect measurements for the six different perovskites. Here, "CsFA" refers to $Cs_{0.15}FA_{0.85}Pb(I_{0.75}Br_{0.25})_3$ and "TC" stands for Triple Cation perovskites $Cs_{0.05}(FA_xMA_y)_{0.95}Pb(I_xBr_y)_3$ with different percentages of formamidinium lead iodide $FAPbI_3$ (*x*) and methylammonium lead bromide $MAPbBr_3$ (*y*), displayed in the graphs as TC *x:y*. **b**, Conductivity values extracted from the three described methods: four contacts in the van der Pauw configuration (4c), average of the two two-opposite contacts (2c) and transfer length method (TLM). **c**, calculated mobilities using the expression $\mu = \sigma_{4c}/(en_0)$ where $\sigma_{4c}$ is the conductivity obtained using the four contacts in the van der Pauw layout and showed in **b**.

**Figure 2a** shows the measured Hall carrier density for the studied compositions. Although the values are different for the different perovskites, the free electron densities in the dark $n_0$ are in the range of $5\times10^{11}$ - $3\times10^{13}$ cm$^{-3}$. We assign this carrier density to the doping-induced charge. The corresponding value for MAPI ($1.9\times10^{12}$ cm$^{-3}$) is in good agreement with results of Hall effect measurements reported in literature.[18] The sample conductivity in the dark σ was assessed by three different measurements. On the same films intended for Hall effect, a voltage was applied to two opposite-located contacts and the current is measured. The results are displayed in **Figure 2b** with the label "2c". The second approach relies on the use of the four contacts at a time, i.e. the voltage is applied to two neighbouring contacts and current is collected in the remaining two. The outcome is also shown in **Figure 2b** under the label "4c".

Finally, it is also possible to estimate the electrical conductivity of a neat perovskite film using the so called transfer length method.[36] A transistor-like structure is used to disentangle the effect of the contact resistance. The working principle, measurement scheme and device architecture are discussed in the **Supporting Information Note S1**. All the perovskite compositions were analysed in the same way and the obtained conductivities are shown in **Figure 2b** under the label of "TLM". **Figure 2b** shows that the transfer length method gives lower values of the conductivity for all the compositions in comparison with the other two approaches.

Hereafter, the mobility $\mu$ is calculated from the conductivity and the carrier density using the equation $\mu = \sigma/(en_0)$ where $e$ is the elementary electronic charge. The results are shown in **Figure 2c** for the different compositions and they lie between 0.5 and 2.5 cm² V⁻¹ s⁻¹. In this case, the conductivities used for the calculations are those determined by the four-contact approach, i.e. $\sigma = \sigma_{4c}$. As $\sigma_{4c}$ is the highest value among the three conductivity sets, the mobilities calculated using the other conductivity sets would be lower. The numerical values of the quantities represented in **Figure 2** are listed in **Supporting Information Table S1**, including the calculated intrinsic carrier concentration $n_i$ for every composition. As the conductivity and the carrier density were measure independently and are linearly related, it is expected that the compositions with a high carrier density, have a higher electrical conductivity. This can be seen in **Figure 2a** and **b** and it holds for the conductivities 4c and 2c. Therefore, normalising these conductivities and the carrier densities to those of one of the compositions, yields a consistent trend which is shown in the **Supporting Information Figure S4**.

From the results of Hall effect measurements, we can estimate that the density of doping induced charge is in the range of 1x10¹¹ to 4×10¹³ cm⁻³. This means that doping density is still far below the critical threshold of the electrode charge ($CV_{bi}$) per cell's volume (approx. 10¹⁶ cm⁻³) as discussed above. Likewise, assuming a $V_{bi} = 1$ V and calculating the depletion width as discussed by Kirchartz, et al.,[48] a depletion width of about 10 µm is obtained, exceeding the actual active layer thickness of 0.5 µm. The determined mobilities are between 0.5 and 2.5 cm² V⁻¹ s⁻¹ which is in good agreement with values found in literature that were obtained with the same approach.[15,18,40]

Intensity-dependent photoluminescence quantum yield

As the recombination of the charge-carriers is influenced by doping, it is possible to indirectly quantify the doping density by means of TRPL and intensity-dependent PLQY measurements.[42,52,53] The total recombination rate $R_{tot}$ can be written as $R_{tot} = -\mathrm{d}(\Delta n)/\mathrm{d}t = k_2 \Delta n(\Delta n + n_0) + k_1 \Delta n$, where $\Delta n$ is the photogenerated (excess) carrier density, and $k_2$ and $k_1$ are the radiative and non-radiative coefficients, respectively. The Auger coefficient has been excluded as it plays a minor role under 1 sun equivalent illumination conditions, and a donor doping density $n_0$ has been introduced.[52,54] The external PLQY is given by[55] $\mathrm{PLQY}_{ext} = J_{rad}/J_{tot} = J_{rad}/(J_{rad} + J_{non\text{-}rad})$, where, in terms of their corresponding recombination rates, the radiative and non-radiative recombination current densities are $J_{rad} = edR_{rad}$ and $J_{non\text{-}rad} = edR_{non-rad}$, respectively. Hence,

$$\mathrm{PLQY}_{ext} = \frac{k_2 \Delta n(\Delta n + n_0)}{k_2 \Delta n(\Delta n + n_0) + k_1 \Delta n} \tag{1}$$

Equation (1) can be plotted as a function of $\Delta n$, which in turn, is related to the intensity (in suns equivalent) by the following expression

$$k_2 \Delta n(\Delta n + n_0) + k_1 \Delta n = \frac{J_{gen}}{ed} = \frac{(\mathrm{suns})J_{SC}}{ed} \tag{2}$$

where $J_{gen}$ is the generation current density, and $J_{SC}$ is the implied short-circuit current density. From the PLQY values as a function of the intensity, the total carrier density $n$ can be calculated as follows

$$n^2 = n_i^2 \exp\left(\frac{\text{QFLS}}{k_B T}\right) \qquad (3)$$

where $k_B$ is the Boltzmann constant, $T$ is the absolute temperature and the quasi-Fermi level splitting (QFLS) is determined as

$$\text{QFLS} = k_B T \ln\left(\text{PLQY} \frac{J_{gen}}{J_{0,rad}}\right) \qquad (4)$$

Here, $J_{0,rad}$ is the maximum recombination current density in the dark and $J_{gen}$ depends on the intensity via equation (2). The analytical expression (1) is plotted in **Figure 3a** as a function of the illumination intensity and the corresponding total carrier density, from equation (3), is shown in **Figure3c.**

As the light intensity, and thereby the generated carrier concentration, decreases ($\Delta n \ll n_0$), the PLQY levels off at different values depending on the doping density $n_0$. In this regime, first-order recombination dominates and consequently the external PLQY does not longer depend on the charge carrier density. From a physical point of view, this levelling out of the PLQY at low intensities is because one of the Fermi levels is pinned by the doping concentration and therefore only one of the Fermi levels shifts with the photogenerated carrier concentration. Above $\Delta n = n_0$, both Fermi levels shift with increasing carrier concentration, which leads to a steeper increase of the QFLS and therefore of the PLQY. **Figure 3a** further shows that as the doping density increases, the PLQY levels out at higher intensities. The same trend against the carrier density is shown in the supporting information of ref.[41] The recombination coefficients used in the **Figure 3a** and **c**, were previously quantified for a triple cation perovskite Cs$_{0.05}$(FA$_{0.83}$MA$_{0.17}$)$_{0.95}$Pb(I$_{0.83}$Br$_{0.17}$)$_3$ and they are $k_1$ = 2×10$^6$ s$^{-1}$ and $k_2$ = 3×10$^{-11}$ cm$^3$ s$^{-1}$ (ref.[34]). Similar trends can be demonstrated for MAPI, where recombination coefficients $k_1$ = 5×10$^6$ s$^{-1}$ and a $k_2$ = 8.1×10$^{-11}$ cm$^3$ s$^{-1}$ (ref.[56]) were used.

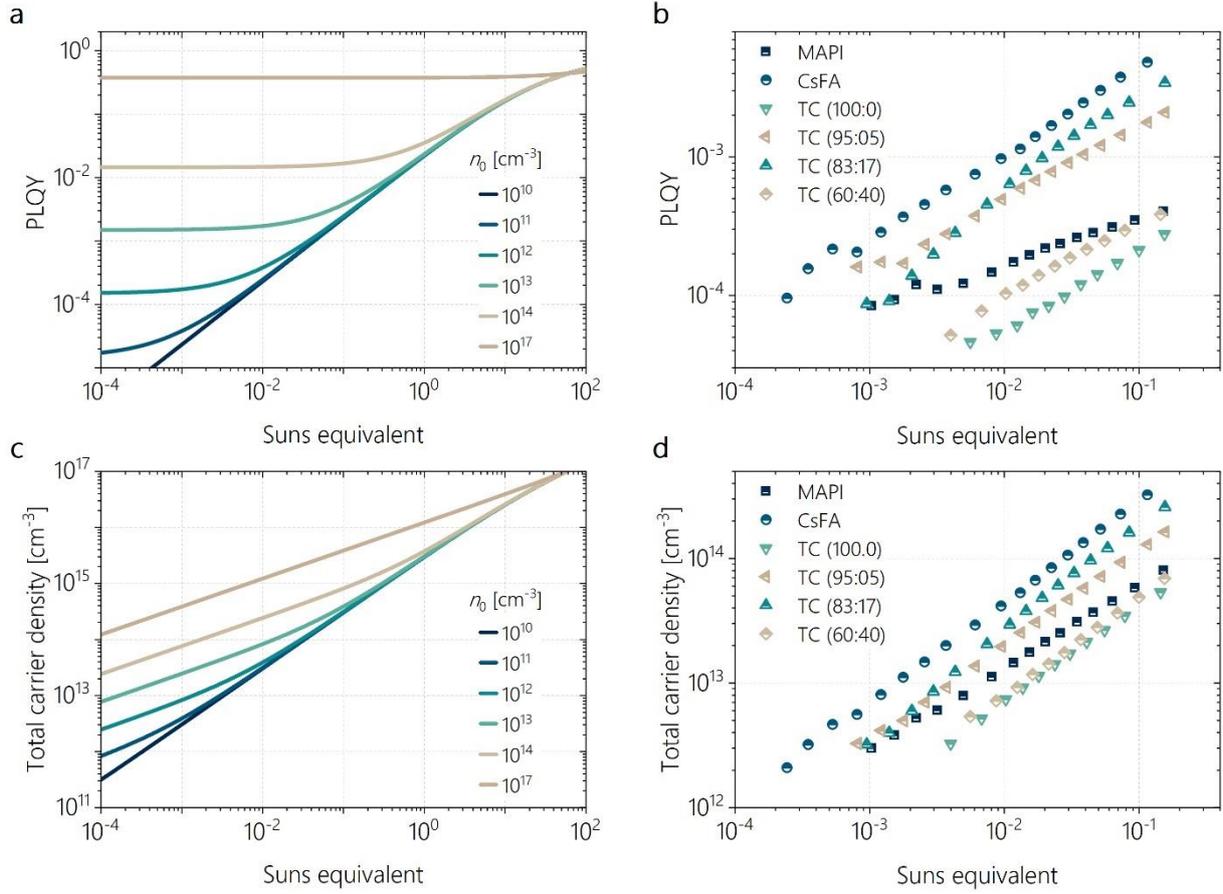

**Figure 3**. **a**, External PLQY as a function of the light intensity in suns equivalent, for different doping densities $n_0$ according to the model of the equation (1) with the constants $k_1 = 2\times10^6$ s$^{-1}$ and $k_2 = 3\times10^{-11}$ cm$^3$ s$^{-1}$ (ref [34]) for a triple cation perovskite Cs$_{0.05}$(FA$_{0.83}$MA$_{0.17}$)$_{0.95}$Pb(I$_{0.83}$Br$_{0.17}$). **b**, Experimental data for the studied compositions, where the light intensity was changed from low to high. **c**, total carrier density for Cs$_{0.05}$(FA$_{0.83}$MA$_{0.17}$)$_{0.95}$Pb(I$_{0.83}$Br$_{0.17}$)$_3$ for different doping densities calculated using the equation (3) and the same constants used for **a**. **d**, Experimental total carrier density $n$ computed using the expression (3) and the data from **b**.

Experimentally, the PLQY can be measured for different light intensities. A neat perovskite film, initially kept in the dark, is illuminated with a 520 nm-wavelength-laser whose intensity can be systematically varied by means of a filter wheel. **Figure 3b** shows the experimental results for the different perovskite films where the light intensity was varied from low to high. Additionally, the trend of the total carrier density against the intensity is shown in **Figure 3d** calculated using the equation (3) and the data in **Figure 3b**. The total carrier density $n$ is of the order of 10$^{15}$ cm$^{-3}$ at 1 sun which matches very well with the results reported in the literature.[22]

As the PLQY depends on the emissive properties of the material, PLQY values below 10$^{-5}$ are already in the detection limit of our setup, therefore, the number of experimental points is different for the range of perovskite compositions. According to **Figure 3d**, this sensitivity limit corresponds to an excess carrier density of 2×10$^{12}$ cm$^{-3}$. Importantly, it is seen that the PLQY does not flatten off for any of the samples at intensities lower that 5×10$^{-3}$ suns.[41] To give an example, **Figure 3b** shows that at 10$^{-3}$ suns equivalent, the PLQY is close to 10$^{-4}$ for the triple cation perovskite Cs$_{0.05}$(FA$_{0.83}$MA$_{0.17}$)$_{0.95}$Pb(I$_{0.83}$Br$_{0.17}$)$_3$. Therefore, according to **Figure 3a**, the implied doping density would be between $10^{11}$ and $10^{12}$ cm$^{-3}$. Hence, from these measurements on these samples, $10^{12}$ cm$^{-3}$ constitutes an upper limit for the doping density in all cases. This outcome is consistent with the electrical conductivity and Hall effect measurements.

Time-resolved photoluminescence amplitude

For continuous illumination, the carrier concentration can be estimated by equation (3) or (4) based on the charge carrier lifetime or the intrinsic carrier concentration. In contrast, photoexcitation by short laser pulses (150 ps in this work) allows direct control of the initially generated charge carrier concentration $\Delta n$. This induced carrier concentration can be calculated from the absorbed photon density per pulse and the thickness of the thin film.

The emitted photoluminescence photon flux is given for relevant conditions ($\Delta n > n_\text{i}$) by equation (5), which depends on $k_2$, $\Delta n$ and $n_0$ (ref [52]).

$$\Phi_\text{PL} \approx k_2[n_0 + \Delta n]\Delta n \qquad (5)$$

Previously, TRPL decays of lead halide perovskites were analysed to determine the doping concentration by F. Staub, *et al.*[42]. However, the TRPL decay can be relatively complex, as it is influenced by charge carrier recombination, separation, diffusion, and extraction. To avoid modelling these processes, here, the initial TRPL amplitude rather than the decay will be analysed using equation (5) for the doping concentration.

**Figure 4a** and **4b** show the TRPL transients for a Sn-based FA$_{0.83}$Cs$_{0.17}$SnI$_3$ and for a Pb-based triple cation Cs$_{0.05}$(FA$_{0.83}$MA$_{0.17}$)$_{0.95}$Pb(I$_{0.83}$Br$_{0.17}$)$_3$ perovskite thin films as function of the carrier concentration, respectively. **Figure 4c** shows that, as far as pure Sn-based perovskite is concerned, the initial TRPL amplitude depends linearly on the carrier concentration and consequently, the doping is larger than the highest induced carrier concentration of 1.4×10$^{16}$ cm$^{-3}$. Regarding the Pb-based, triple cation perovskite Cs$_{0.05}$(FA$_{0.83}$MA$_{0.17}$)$_{0.95}$Pb(I$_{0.83}$Br$_{0.17}$)$_3$, the initial fast decay in **Figure 4b** becomes more pronounced with increasing intensity which means, according to the corresponding trend in **Figure 4c**, that the amplitude increases with the square of the induced carrier concentration. As a result, the doping is smaller than the smallest induced carrier concentration of 4×10$^{13}$ cm$^{-3}$. Assuming similar external radiative coefficients $k_2$ in both materials, the doping density for the pure Sn-based perovskite is about 10$^{17}$ cm$^{-3}$, indicated by the vertical arrow in **Figure 4c**.

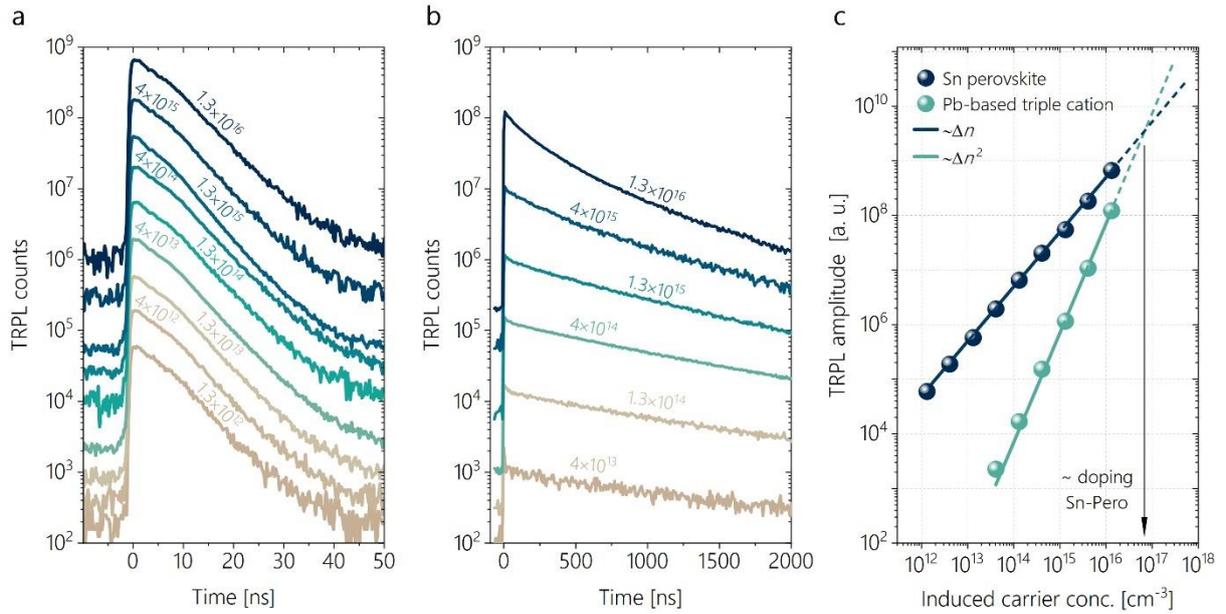

**Figure 4. a** and **b,** Injection dependent TRPL transients for a tin-based FA$_{0.83}$Cs$_{0.17}$SnI$_3$ perovskite and a lead-based triple cation perovskite Cs$_{0.05}$(FA$_{0.83}$MA$_{0.17}$)$_{0.95}$Pb(I$_{0.83}$Br$_{0.17}$)$_3$, respectively, as function of the injected carrier concentration. **c,** TRPL amplitude as function of injected carrier concentration corresponding to the same films on **a** and **b**.

## Dark CELIV

While the measurement presented above were used to estimate the doping density in neat perovskite films, in the following, we aim to quantify the doping density in device configuration. To this end, we performed (CELIV) measurements. CELIV is a powerful technique that allows the user to measure doping densities as well as mobile ion densities. Using an additional pulsed laser excitation (photo-CELIV), it is also possible to assess the photogenerated carrier density. Every transient directly shows the presence of additional charges if they are present in relevant conditions with regards to the $CU_{\text{max}}$ charge per unit volume.[47,57,58] The latter can be controlled by adjusting the applied bias $U_{\text{max}}$ to the metal electrodes.[32]

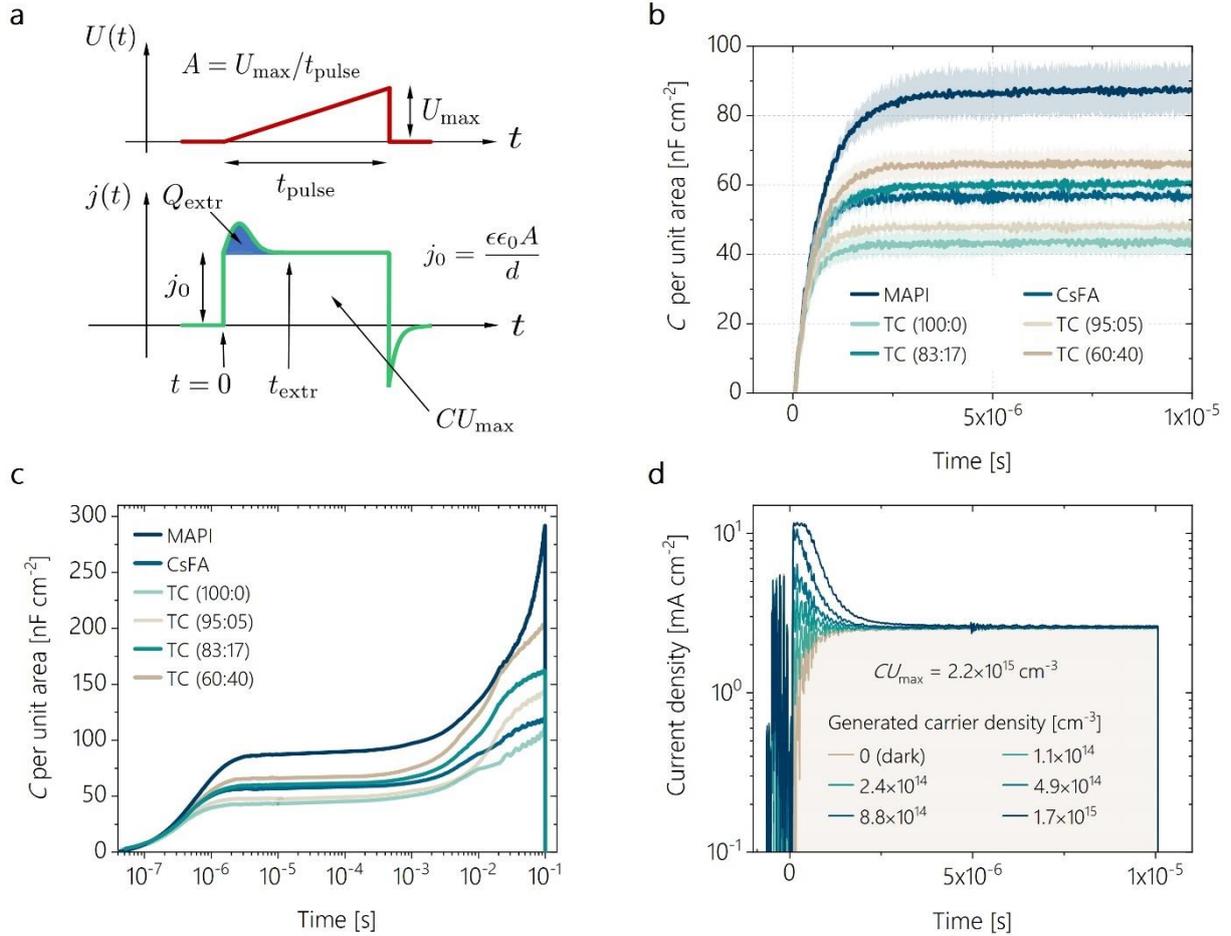

**Figure 5. a**, Schematics of the CELIV technique. The CELIV voltage ramp is applied during a time $t = t_{\text{pulse}}$ while the measured current is recorded with an oscilloscope. In case a free electronic charge concentration larger than the $CV_{\text{bi}}$ charge is present, a bump on top of the $CV_{\text{bi}}$ can be observed during the initial phase of the extraction. The integrated area of such a bump corresponds to the concentration of free electronic charges. **b**, Dark-CELIV transients for the different investigated perovskite compositions on the microsecond timescale. **c**, Complete dark-CELIV transients extending to slower timescales, showing the presence of mobile ions. **d**, Photo-CELIV transients for a Cs$_{0.05}$(FA$_{0.83}$MA$_{0.17}$)$_{0.95}$Pb(I$_{0.83}$Br$_{0.17}$)$_3$ triple cation perovskite showing the photogenerated charge after the laser pulse excitation. The shaded area represents the $CU_{\text{max}}$ charge which amounts to 2.2×10$^{15}$ cm$^{-3}$. The plot mainly demonstrates that if a large enough concentration of electronic charges were present, they would show up as a bump on top of the $CU_{\text{max}}$ charge.

In this measurement, the cell is reverse-biased by a linearly increasing voltage with amplitude $U_{\text{max}}$, as it is shown in **Figure 5a**. As a result of the accumulation of charges on the contacts of the solar cell, a displacement current density $j_0$ builds up in the *RC* circuit, which is subsequently measured with an oscilloscope. Previous experiments[47,59,60] have shown that electronic charges are extracted from the device on microsecond timescales. Therefore, any doping-induced charges that are present in the active layer should be extracted at the same timescales. Their presence will be visible only if the doping-induced charge density is higher than the $CU_{\text{max}}$ charge per cell volume (or $CV_{\text{bi}}$ if $CU_{\text{max}} = CV_{\text{bi}}$). If this is the case, a 'bump' will appear on top of the $CU_{\text{max}}$ charge, as depicted in **Figure 5a**. The integrated area under this 'bump' corresponds to the doping-induced charge whose volume density is $Q/e$.

As it is shown in **Figure 5b**, none of the displayed dark-CELIV transients show the 'bump' on the microsecond timescales. Therefore, the free carriers that might be present in the active layer are not enough to make an appearance above the $CU_\mathrm{max}$ charge. However, as $t_\mathrm{pulse}$ increases because the voltage is ramped more slowly, a rise beyond the $CU_\mathrm{max}$ charge is observed at milliseconds to seconds timescales. This effect is shown in the **Figure 5c** for the range of studied compositions (the raw transients are provided in **Supporting Information Figure S7**). Such a current increase at longer timescales has previously been linked to the presence of mobile ions,[27,30,61–63] which have a much lower mobility than electronic charges. Therefore, performing CELIV measurements at longer timescales allows us to estimate the density and diffusion constant of the effective ionic concentration, as previously shown by Thiesbrummel *et al.* for PbSn perovskite systems.[34]

To further prove that the CELIV technique is able to estimate doping densities in relation to the $CU_\mathrm{max}$ charge per unit volume, we used photo-CELIV.[47] In this experiment, a laser pulse is used to generate free charges in the device. The laser pulse is timed such that it excites these charges right when the voltage ramp starts. Using different laser fluences, a range of known free carrier concentrations can be excited in the bulk of the material. This allows us to prove that if there were enough charges in the active layer, we would in fact measure them and hence the fact that we did not see a bump in the dark means that there are not enough charges. The result can be seen in **Figure 5d** for different light intensities (expressed as total induced carrier density). The data displayed here was measured on a triple cation perovskite cell $Cs_{0.05}(FA_{0.83}MA_{0.17})_{0.95}Pb(I_{0.83}Br_{0.17})_3$. The transients demonstrate that when the generated charge carrier density is below $10^{14}$ cm$^{-3}$, their detection is hindered by the presence of the $CU_\mathrm{max}$ charge. In other words, generated free-charge-carrier-densities below $10^{14}$ cm$^{-3}$ are buried below the $CU_\mathrm{max}$ charge.

Although the doping density in equilibrium is significantly below the required threshold ($CU_\mathrm{max}$) to influence the device performance, mobile ion density is comparable or exceeds the $CV_\mathrm{bi}$ charge per unit volume. Therefore, although perovskites cannot be considered to have a performance-relevant doping density, mobile ions are present in concentrations large enough to form field-free regions and junctions in the devices under steady-state conditions, thereby impacting device performance.

## **Conclusions**

Through an extensive experimental investigation, comprised of different optical and electrical experimental techniques, we collected evidence to address the controversial question of whether perovskite thin films behave like doped semiconductors. By introducing the concept of the electrode charge, we derived a critical threshold at which doping matters, which is supported by drift-diffusion simulations. For a broad range of perovskite compositions, Hall effect measurements, carried out using an AC field and lock-in detection, showed that the density of doping induced charge is between $5\times10^{11}$ - $3\times10^{13}$ cm$^{-3}$. For MAPI, this is in good agreement with earlier observations ($1.9\times10^{12}$ cm$^{-3}$) using Hall effect measurements. The electrical conductivity of the studied compositions was measured in three different ways, namely two-contacts, four-contacts and transfer length method. The carrier mobility was calculated from the carrier density determined by Hall effect and the four-contact conductivity. The values are between 0.5 and 2 cm$^2$ V$^{-1}$ s$^{-1}$ which are in good agreement with those values found in literature for similar perovskite systems. We then measured the intensity-dependent PLQY and TRPL amplitudes. By these two methods, upper limits for the doping induced charge density of 10$^{12}$ cm$^{-3}$ and of 5×10$^{13}$ cm$^{-3}$, were deduced for Pb-based perovskites, respectively. Consequently, these results are consistent with the carrier density determined by Hall effect. While the above-mentioned measurements were conducted on bare perovskite films, we additionally used CELIV measurements to determine the doping density in full devices. The results obtained show that the doping density is lower than the charge carrier density on the electrodes $CU_\mathrm{max}$ which amounts to $2.2 \times 10^{15}$ cm$^{-3}$. The latter value is in turn less than the critical threshold for the electrode charge density $CV_\mathrm{bi}$ per cell volume, which is of the order of $10^{16}$ cm$^{-3}$. The consistency of these measurements allows us to conclude that the doping density is negligible for the device performance for most Pb-based perovskites. Nonetheless, for halide perovskites incorporating other metals, which are known to easily oxidise and self-dope, doping is more likely to impact device performance. Considering that for a *p-n* junction to be formed, the minimum doping densities are on the order of 10$^{16}$ cm$^{-3}$, we are able to conclude that the doping densities of the Pb-based halide perovskites investigated in this work are not high enough to influence the internal band bending. Finally, this study also shows that although the doping density is low, the mobile ion density does exceed critical $CV_\mathrm{bi}$ threshold, thereby creating field-free regions in the device and apparent junctions.

### **Acknowledgments**

We acknowledge funding from the Deutsche Forschungsgemeinschaft (DFG, German Research Foundation) within the SPP 2196 (SURPRISE 423749265 and HIPSTER 424709669). We further acknowledge financial support by the Federal Ministry for Economic Affairs and Energy within the framework of the 7th Energy Research Programme (P3T-HOPE, 03EE1017C) and HyPerCells (a joint graduate school of the Potsdam University and the Helmholtz-Zentrum Berlin für Materialien und Energie). F.L. acknowledges financial support from the Alexander von Humboldt Foundation via the Feodor Lynen program.

### **Data availability statement**

The data that support the findings of this study are available from the corresponding author upon reasonable request.

# Revealing the doping density in perovskite solar cells and its impact on device performance

Francisco Peña-Camargo[1], Jarla Thiesbrummel[1,2], Hannes Hempel[3], Artem Musiienko[4], Vincent M. Le Corre[1,5], Jonas Diekmann[1], Jonathan Warby[1], Thomas Unold[3], Felix Lang[1], Dieter Neher[1], Martin Stolterfoht[1,*]

1. Physik weicher Materie, Institut für Physik und Astronomie, Universität Potsdam, Karl-Liebknecht-Str. 24–25, 14776 Potsdam, Germany
2. Clarendon Laboratory, University of Oxford, Parks Road, Oxford OX1 3PU, United Kingdom
3. Department of Structure and Dynamics of Energy Materials, Helmholtz-Zentrum Berlin für Materialien und Energie, D-14109, Berlin, Germany
4. Institut für Silizium-Photovoltaik, Helmholtz-Zentrum Berlin für Materialien und Energie, Kekuléstrasse 5, 12489, Berlin, Germany
5. Institute of Materials for Electronics and Energy Technology, Friedrich-Alexander-Universität Erlangen-Nürnberg, 91058, Erlangen, Germany

email: stolterf@uni-potsdam.de

## Methods

### Device Fabrication of *pin*-type cells:

*Substrates and HTL:*

Pre-patterned 2.5x2.5 cm$^2$ 15 Ω/sq. ITO substrates (Psiotec, United Kingdom) were cleaned with acetone, 3% Hellmanex solution, deionised water and iso-propanol, by sonication for 10 minutes in each solution. After an oxygen plasma treatment (4 min, 120 W), the samples were transferred to a N$_2$-filled glovebox. For the *pin*-type cells shown in the main text, a PTAA (Sigma-Aldrich) layer with thickness of 8 nm was spin-coated from a 1.75 mg ml$^{-1}$ PTAA/toluene solution at 6000 rpm for 30 seconds. After 10 min annealing on a hotplate at 100 °C, the films were cooled down to room temperature and a 60 μL solution of PFN-Br (1-Material, 0.5 mg/mL in methanol) was deposited onto PTAA dynamically at 4000 rpm for 30 s resulting in a film with thickness below the detection limit of our AFM (< 5 nm). No further annealing occurred.

*Perovskite solutions:*

The solutions for the lead-based perovskites were prepared as follows: 1.2 M FAPbI$_3$ solution was prepared by dissolving FAI and PbI$_2$ in DMF:DMSO (4:1 volume ratio) which contains a 10 molar % excess of PbI$_2$. The 1.2 M MAPbBr$_3$ solution was made by dissolving MABr and PbBr$_2$ in DMF:DMSO (4:1 volume ratio) which contains a 10 molar % excess of PbBr$_2$. The solutions were stirred overnight at room temperature. By mixing these FAPbI$_3$ and MAPbBr$_3$ solutions in a volume ratio of *x:y*, we get what we call "MAFA" solutions. Lastly, 42 μL of a 1.5 M CsI solution in DMSO was mixed with 958 μL of each one of the MAFA solutions resulting in nominal triple cation perovskite stoichiometries of Cs$_{0.05}$(FA$_x$MA$_y$)$_{0.95}$Pb(I$_x$Br$_y$)$_3$. These triple cation perovskites are named "TC *x:y* (%)" in the manuscript. The MAPI solution was prepared by dissolution of MAI powder (Dysol 1.3 M) with PbI$_2$ (TCI, 1.3 M) in a γ-butyrolactone/DMSO mixed solvent (7:3 by volume), while stirring at 60 °C for 10 min.

The 1.2 M tin-based FA$_{0.83}$Cs$_{0.17}$SnI$_3$ perovskite solution was prepared by dissolving FAI, CsI and SnI$_2$, together with 10 molar % SnF$_2$, in a 4:1 DMF:DMSO mixture. The nominal stoichiometry of the compound is. We note that all precursors used for this solution were stored and weighed in a N$_2$-filled glovebox, to prevent contamination of the solution with O$_2$ or H$_2$O. The solution was stirred for 2 hours at room temperature. Finally, the solution was filtered using a 0.45 μm PTFE filter.

*Perovskite film fabrication*:

All triple cation perovskite films were deposited by spin-coating at 4000 rpm for 40 s and 10 s after the start of the spinning process, the spinning substrate was washed with 300 μL ethyl acetate for approximately 1 s (the anti-solvent was placed in the centre of the film). We note, that by the end of the spinning process the perovskite film turned dark brown. The perovskite film was then annealed at 100 °C for 1 hour on a preheated hotplate where the film turned slightly darker.

The tin-based FA$_{0.83}$Cs$_{0.17}$SnI$_3$ perovskite films were deposited by spin-coating at 3000 rpm for 45 s with a ramp of 1000 rpm/s. 25 s after the start of the spinning process, the spinning substrate was washed with 200 μL anisole, which was deposited in the centre of the film. By the end of the spinning process, the perovskite films turned dark brown. The perovskite films were then annealed at 100 °C for 10 minutes on a preheated hotplate inside the $N_2$ filled glovebox. During the annealing process, the perovskite films turned black.

*ETL and Top Contact*:

After annealing, the samples were transferred to an evaporation chamber where fullerene-$C_{60}$ (30 nm), 2,9-Dimethyl-4,7-diphenyl-1,10-phenanthroline BCP (8 nm) and copper (100 nm) were deposited under vacuum (p = $10^{-7}$ mbar). The overlap of the copper and the ITO electrodes defined the active area of the pixel (6 mm$^2$).

**Photoluminescence Measurements:** Excitation for the PL imaging measurements was performed with a 520 nm CW laser (Insaneware) through an optical fibre into an integrating sphere. The intensity of the laser was adjusted to a 1 sun equivalent intensity by illuminating a 1 cm$^2$-size perovskite solar cell under short-circuit and matching the current density to the $J_{SC}$ under the sun simulator (e.g. ~22.0 mA/cm$^2$ at 100 mWcm$^{-2}$, or 1.375x10$^{21}$ photons m$^{-2}$s$^{-1}$ for a 83-17 triple cation perovskite cell). A second optical fibre was used from the output of the integrating sphere to an Andor SR393i-B spectrometer equipped with a silicon CCD camera (DU420A-BR-DD, iDus). The system was calibrated by using a halogen lamp with known spectral irradiance, which was shone into to integrating sphere. A spectral correction factor was established to match the spectral output of the detector to the calibrated spectral irradiance of the lamp. The spectral photon density was obtained from the corrected detector signal (spectral irradiance) by division through the photon energy ($h\nu$), and the photon numbers of the excitation and emission were obtained from numerical integration using Matlab. In a last step, three fluorescent test samples with high specified PLQY (~70%) supplied from Hamamatsu Photonics where measured where the specified value could be accurately reproduced within a small relative error of less than 5%.

**Intensity-dependent PL Measurements:** The samples were illuminated in the integrating sphere as described above. A continuously variable neutral density filter wheel (ThorLabs) was used to attenuate the laser power to measure at different intensities which was monitored using an additional Si photodetector. The samples were illuminated at a given intensity for a variable illumination time using an electrical shutter. The illumination time was set to 1 second for the measurements shown in the main text. After the specified illumination the PL spectra were recorded by averaging 100 spectra taken using a detector exposure time of 30 μs. Then the electrical shutter was closed, the filter wheel was moved to the next position and the steps were repeated every 10 degrees. A custom-built Labview code was written to automate the measurement, and a Matlab code to automate the data evaluation.

**Current density-voltage characteristics:** $JV$-curves were obtained in a 2-wire source-sense configuration with a Keithley 2400. An Oriel class AAA Xenon lamp-based sun simulator was used for illumination providing approximately 100 mW cm$^{-2}$ of AM1.5G irradiation and the intensity was monitored simultaneously with a Si photodiode. The exact illumination intensity was used for efficiency calculations, and the simulator was calibrated with a KG5 filtered silicon solar cell (certified by Fraunhofer ISE). The obtained short-circuit current density ($J_{SC}$) is checked by integrating the product of the External Quantum Efficiency and the solar spectrum which matches the obtained $J_{SC}$ within less than 5%. The temperature of the cell was fixed to 25 °C and a voltage ramp (scan rate) of 67 mV/s was used.

**Transient PL:** TrPL measurements shown in the main text were carried out in a home-built setup using 660 nm excitation laser light from a supercontinuum light source (SuperK) with a 25-35 μm spot size. We chose the longer wavelength excitation to avoid effects of charge diffusion from a high to low carrier density region. The excitation pulse had a repetition rate of 150 kHz and the PL emission was collected panchromatically through a photomultiplier and time-correlated single photon counting technique. The fluence was controlled with a tuneable neutral density filter and monitored with a power meter.

**Charge extraction by linearly increasing voltage (dark-CELIV):** In dark dark-CELIV the device was initially held at short-circuit conditions. Then the voltage was linearly increased to minus 0.4 V (in reverse bias) using a pulse generator. The slope ($A$) was thereby varied to assess a wide timescale range. The current transients were recorded with an oscilloscope (Agilent DSO9104H) and measured with a variable load resistance (50 Ω at short 10 μs pulse and up to 10 kΩ at 100 ms pulses) to keep the voltage response approximately constant. The increased load resistance reduces the time resolution at short times but allows to record the response for long pulses. The continuous increase of the electrode charge leads to a step-like voltage response. The voltage response step of the cell is given by $\Delta V = AR_{\text{load}}C$ from which we calculated the capacitance of the cell ($C$). Equilibrium charges in the active layer (doping-induced electronic charge or mobile ions) lead to an additional bump in the voltage response as further discussed in **Supporting Information Figure S7**.

**AC Hall measurements:** Hall measurements were performed in 4-probe configuration with Lake Shore 8400 Hall system. All 4 probe combinations showed linear *IV* response and ohmic-contacts signature. We used AC magnetic field (50 mHz) and a lock-in amplifier to enhance the Hall signal. The current through the samples (in the range 0.1-2 nA) was supported by the current source. The Hall data are fully consistent with 2-probe (IV) and 4-probe conductivity measurements see **Supporting Information Figure S4.**

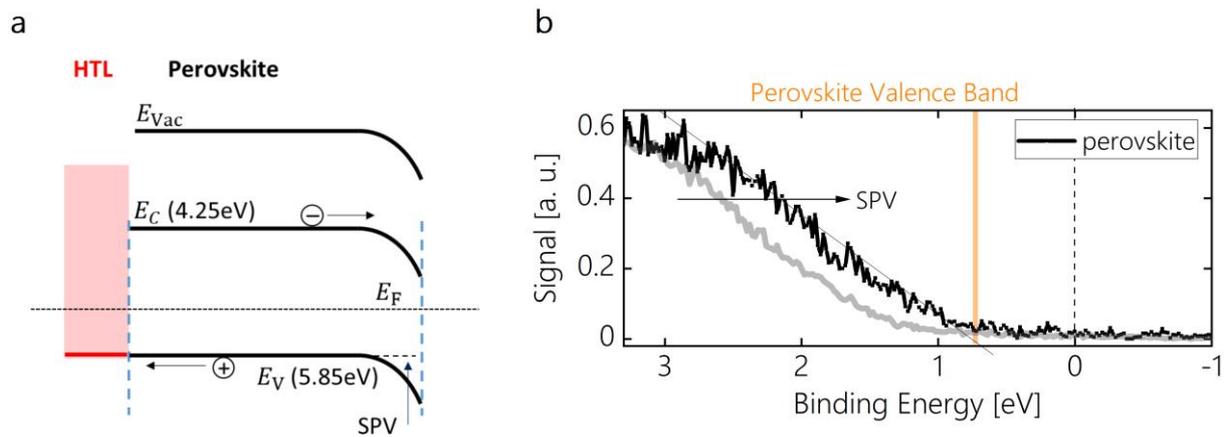

**Figure S1. a**, Sketch of a band diagram for a partial perovskite solar cell stack. The perovskite surface appears to be *n*-doped due to band bending, for example at the surface or towards the hole transport layer (HTL).[22] **b**, The apparent valence band onset is found to be 1.35 eV. However, upon 1 sun equivalent illumination, the bands are flattened at the surface, shifting the valence band onset by the surface photovoltage (SPV). This allows us to assess the valence band onset in the bulk which is approximately 0.8 eV away from the Fermi level, regardless if the band bending is the perovskite surface or the HTL.[22,24]

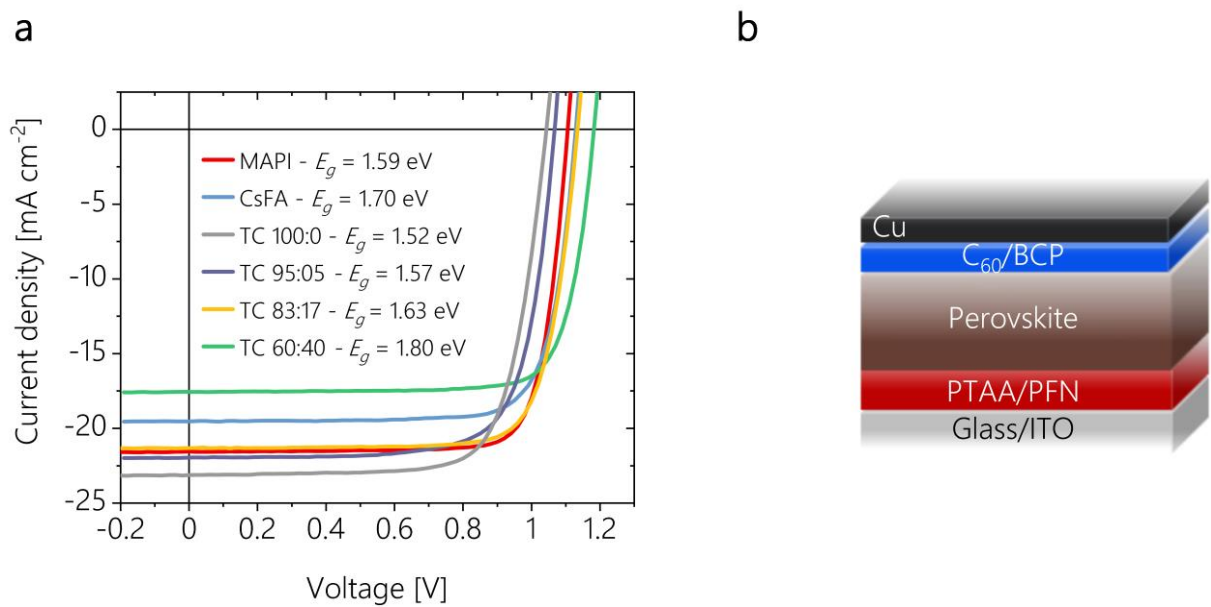

**Figure S2. a**, Current density versus voltage curves of the cells highlighting their bandgap. **b**, Layout of the devices in the inverted *pin*-structure. Here, "CsFA" refers to $Cs_{0.15}FA_{0.85}Pb(I_{0.75}Br_{0.25})_3$ and "TC" stands for Triple Cation perovskites $Cs_{0.05}(FA_xMA_y)_{0.95}Pb(I_xBr_y)_3$ with different percentages of formamidinium lead iodide $FAPbI_3$ ($x$) and methylammonium lead bromide $MAPbBr_3$ ($y$), in such a way that $x + y = 100\%$.

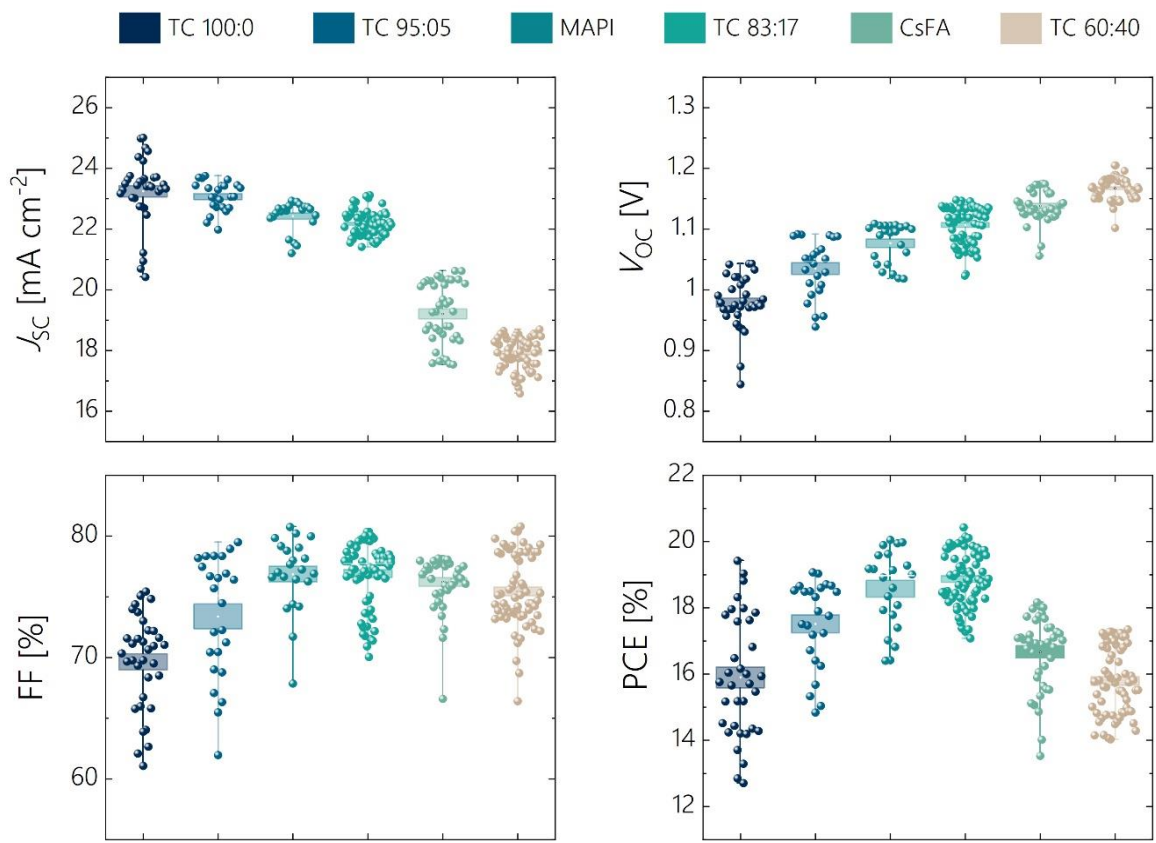

**Figure S3.** Performance parameters of the investigated solar cells. The top labels correspond to the different investigated compositions. Here, "CsFA" refers to $Cs_{0.15}FA_{0.85}Pb(I_{0.75}Br_{0.25})_3$ and "TC" stands for Triple Cation perovskites $Cs_{0.05}(FA_xMA_y)_{0.95}Pb(I_xBr_y)_3$ with different percentages of formamidinium lead iodide $FAPbI_3$ ($x$) and methylammonium lead bromide $MAPbBr_3$ ($y$), in such a way that $x + y = 100\%$.

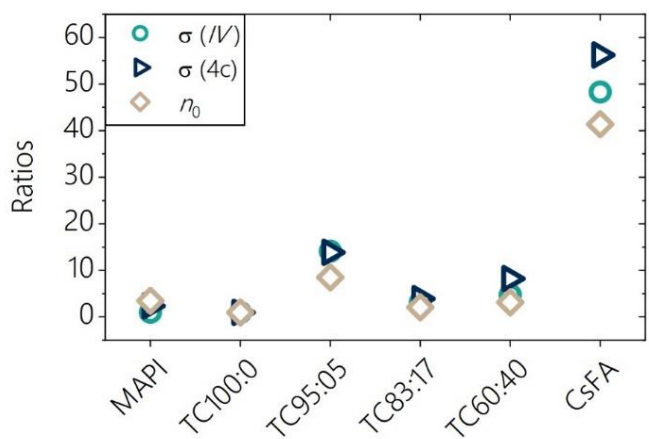

**Figure S4.** Ratios of the independent conductivities and carrier densities to those corresponding to the composition $Cs_{0.05}FA_{0.95}PbI_3$ which exhibits the lower values.

## Note 1

If the field is parallel to the current density (ohmic condition[9,38]), the conductivity can be determined for the different transfer lengths $L_1$ according to the expression

$$\frac{V}{I} = \frac{1}{\sigma}\frac{L_1}{L_2 L_3} + R_C \qquad (S1)$$

where $V$ is the applied bias, $I$ is the measured current, $L_2$ is the thickness of the active layer, and $R_C$ is the contact resistance. In the inset of **Supporting Information Figure S4**, a linear trend is observed.

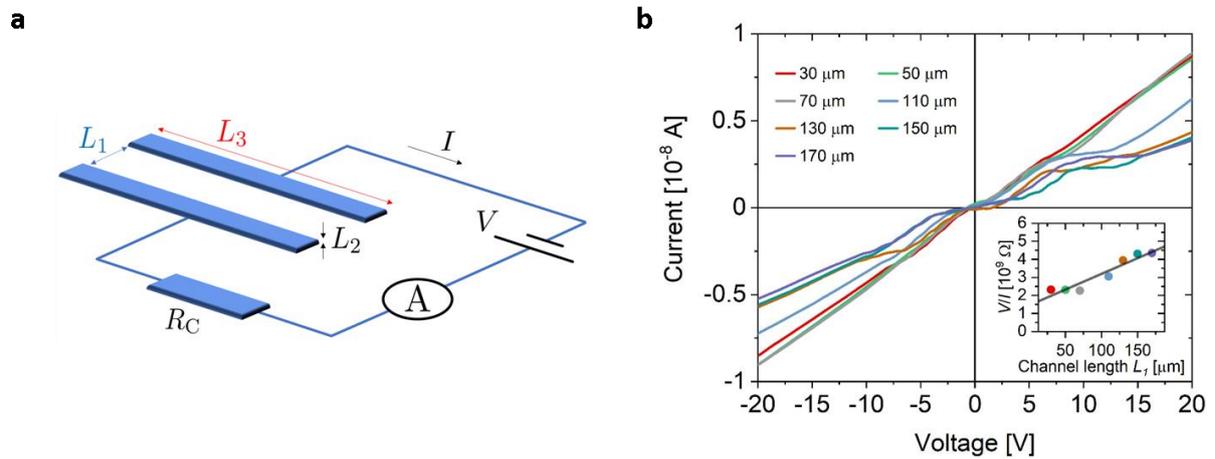

**Figure S5. a**, Scheme of the transfer length method. Here $L_3$=1 cm and $L_1$ ranges from 30 μm to 170 μm, **b**, Example of the measurement of the current as a function of the applied bias for different transfer lengths $L_1$. In the inset, the quotients $V/I$ are shown for several transfer lengths $L_1$. The conductivity is determined from the slope of this line.

| Composition | $n_0$ (Hall) [cm$^{-3}$] | σ (4c) [Ω$^{-1}$ cm$^{-1}$] | σ (2c) [Ω$^{-1}$ cm$^{-1}$] | σ (TLM) [Ω$^{-1}$ cm$^{-1}$] | μ (4c) [cm$^2$ V$^{-1}$ s$^{-1}$] | $n_i$ [cm$^{-3}$] |
|---|---|---|---|---|---|---|
| MAPbI$_3$ | 1.88×10$^{12}$ | 1.85×10$^{-7}$ | 5.51×10$^{-8}$ | 2.20×10$^{-8}$ | 0.62 | 5.0×10$^4$ |
| CsFA | 2.23×10$^{13}$ | 4.35×10$^{-6}$ | 2.65×10$^{-6}$ | 2.80×10$^{-8}$ | 1.22 | 5.6×10$^3$ |
| TC (100:0) | 5.39×10$^{11}$ | 7.73×10$^{-8}$ | 5.47×10$^{-8}$ | 1.30×10$^{-7}$ | 0.90 | 2.0×10$^5$ |
| TC (95:05) | 4.59×10$^{12}$ | 1.07×10$^{-6}$ | 7.74×10$^{-7}$ | 1.30×10$^{-7}$ | 0.82 | 7.3×10$^4$ |
| TC (87:17) | 1.12×10$^{12}$ | 3.06×10$^{-7}$ | 1.73×10$^{-7}$ | 7.80×10$^{-8}$ | 1.71 | 2.2×10$^4$ |
| TC (60:40) | 1.71×10$^{12}$ | 6.37×10$^{-7}$ | 2.62×10$^{-7}$ | 6.20×10$^{-8}$ | 2.02 | 0.8×10$^3$ |

**Table S1**. Numerical values of the data shown in **Figure 1**. Here, "CsFA" refers to Cs$_{0.15}$FA$_{0.85}$Pb(I$_{0.75}$Br$_{0.25}$)$_3$ and "TC" stands for Triple Cation perovskites Cs$_{0.05}$(FA$_x$MA$_y$)$_{0.95}$Pb(I$_x$Br$_y$)$_3$ with different percentages of formamidinium lead iodide FAPbI$_3$ ($x$) and methylammonium lead bromide MAPbBr$_3$ ($y$). The intrinsic carrier concentration $n_i$ was calculated using an effective electron mass of $0.21 m_e$.

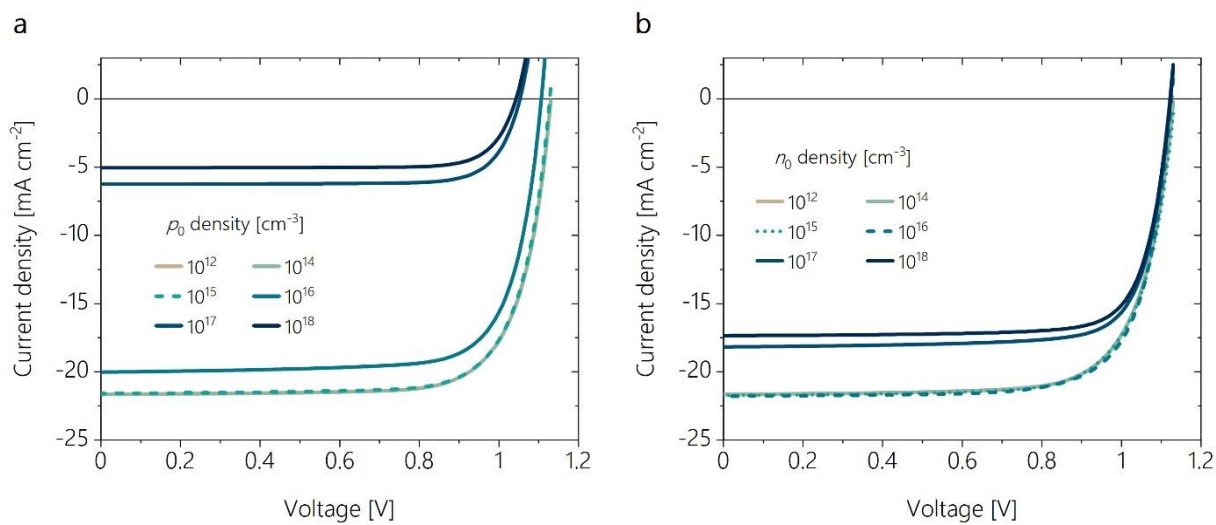

**Figure S6.** Simulated *JV* curves for different doping densities with a mobile cation density of $10^{17}$ cm$^{-3}$. The simulations were carried out in Fluxim-Setfos. **a**, acceptors ($p_0$). **b**, donors ($n_0$).[64]

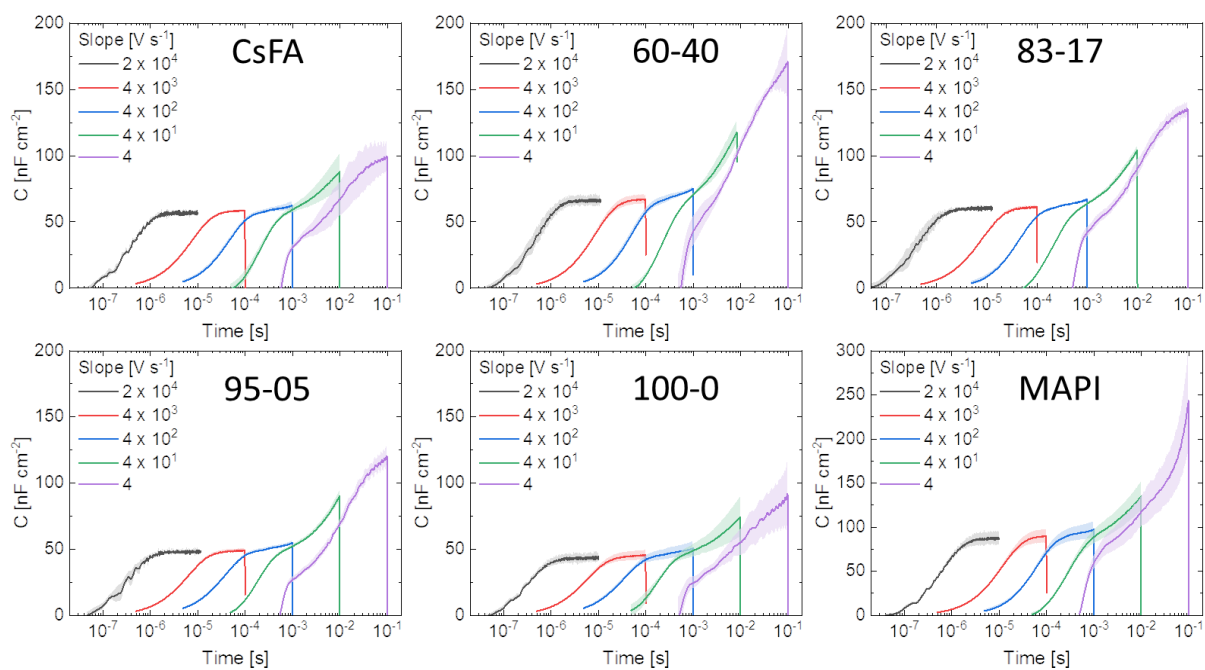

**Figure S7.** Dark CELIV currents for the six different perovskites and several slopes $A=U_{max}/t_{pulse}$ for each one.